\documentclass[prb,twocolumn,amsmath,amssymb,showpacs]{revtex4}
\usepackage{graphicx}

\def\beq{\begin{eqnarray}}
\def\eeq{\end{eqnarray}}
\def\beqa{\begin{eqnarray}}
\def\eeqa{\end{eqnarray}}

\begin{document}

\title{Evidence for two competing order parameters in underdoped cuprates superconductors from a model analysis of the Fermi-arc 
effects
}

\author{Andr\'es Greco}
\affiliation{
Facultad de Ciencias Exactas, Ingenier\'{\i}a y Agrimensura and
Instituto de F\'{\i}sica Rosario
(UNR-CONICET).
Av. Pellegrini 250-2000 Rosario-Argentina.
}

\date{\today}

\begin{abstract}
Preformed pairs above $T_c$ and the two-gap scenarios are two main proposals for describing the low doping pseudogap phase of high-$T_c$ cuprates. Very recent angle-resolved photoemission experiments have shown features which were 
interpreted as evidence for preformed pairs. Here it is shown that those results can be explained also in the context of the two-gap scenario 
if self-energy effects are considered. The discussion is based on the $d$-CDW theory or the flux phase of the $t-J$ model. 
\end{abstract}


\maketitle

The low doping pseudogap (PG) phase 
of high-$T_c$ cuprates is one of the most relevant  and controversial puzzles  
in condensed matter physics.\cite{timusk99} Two main scenarios are disputing the explanation 
of the PG phase. In one of the scenarios the PG is associated to preformed pairs  existing above $T_c$,
while in the other, the so called  two-gap scenario, the superconducting gap and the PG  have different origins and may compete
each other.\cite{hufner08}
Whereas some experiments provide support for preformed pairs\cite{alloul93,fischer07,wang05,wang06} others are interpreted in terms  
of two competing phases.\cite{letacon06,yu08,mook08,liu08,khasanov08,ma08,yoshida09,kondo09} 
One of the most unexpected and surprising results  in the PG phase arises from angle-resolved photoemission spectroscopy (ARPES) which show 
that the Fermi surface (FS) is broken up into disconnected arcs.\cite{norman98,damascelli03,timusk99}
Very recent ARPES experiments which were interpreted in terms of preformed pairs\cite{kanigel06,kanigel07,kanigel08} brought new insights to this 
discussion.
Interestingly, in Ref.[\onlinecite{kanigel08}] it was found that in the PG phase electronic excitations 
near the anti-node disperse like in a superconductor. Then, it is crucial to evaluate  to what extent  
ARPES experiments may be compatible  with the two-gap scenario. 
A well known theory for two competing orders is the $d$-CDW theory\cite{chakravarty01} which proposes,  phenomenologically, a 
normal state complex order parameter with $d$-wave symmetry (see Ref.[\onlinecite{varma06}] for a related proposal). 
Interestingly,
at mean field level the $t-J$ model exhibits, at low doping and below a characteristic temperature $T_{FP}$, the flux phase (FP) 
which possesses the main properties to be identified with the $d$-CDW.\cite{affleck88,kotliar88,morse91,cappelluti99}
In this paper it will be shown that some  ARPES features can be understood  after considering, beyond mean field,  self-energy effects showing that a theory for two competing orders is compatible with  the experiments. 
\begin{figure}
\begin{center}
\setlength{\unitlength}{1cm}
\includegraphics[width=8.8cm,angle=0.]{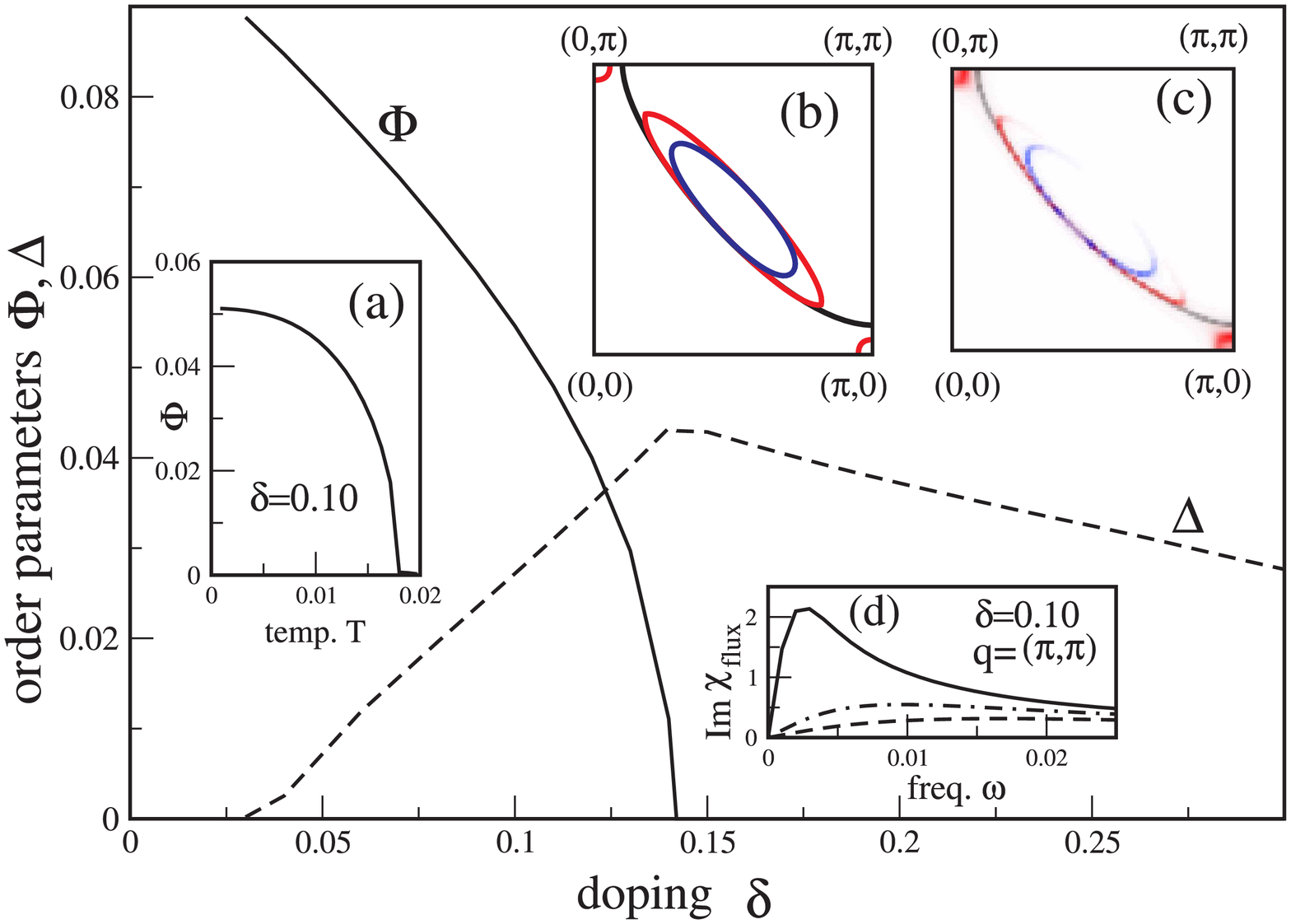}
\end{center}
\caption{ (color online) Doping dependence of the superconducting ($\Delta$) and $d$-CDW ($\Phi$) order parameters at $T=0$.
Inset (a) Temperature dependence of the $d$-CDW order parameter $\Phi$ for $\delta=0.10$.  
Inset (b) Evolution of the FS with decreasing temperature. Black line is the FS for $T>T_{FP}$, red line for $T = 0.017$ ($\Phi \sim 0.01$) 
and blue line for $T = 0.01$ ($\Phi\sim 0.05$).
Inset (c) Spectral weight intensity on the FS for the same temperatures than inset (b), showing 
that the intensity of the outer part of the pockets is low.
Inset (d) Softening of the $d$-wave flux mode approaching $T_{FP}$. $Im \chi_{flux}$ for 
$T=0.030$ (dashed line), $T=0.025$ (dot-dashed line) and $T=0.023$ (solid line). 
}\label{fig1}
\end{figure}

Before presenting results beyond mean field level, the physical picture emerging from the mean field solution of the $t-J$ model is given for completeness. 
The mean field treatment of the $t-J$ model yields a quasi-particle (QP) dispersion $\epsilon_{k}=-2(t \delta+
rJ)(cos(k_x)+cos(k_y))-4t' \delta cos(k_x) cos(k_y)-\mu$, with $r=1/N_s \sum_{\bf k} cos(k_x) n_F(\epsilon_{k})$.\cite{greco04} $n_F$ is the Fermi function, $\delta$ the doping away from half-filling, $\mu$ the chemical potential and $N_s$ the number of sites. 
$t$, $t'$ and $J$ are hopping between nearest neighbor, next nearest neighbor and the nearest neighbor Heisenberg copling respectively.
At low doping, the homogeneous mean field solution becomes unstable against a FP or $d$-CDW with order parameter $\Phi(k)=\Phi \gamma(k)$
which has $d$-wave symmetry like the superconducting gap 
$\Delta(k)=\Delta \gamma(k)$ being $\gamma(k)=1/2(cos(k_x)-cos(k_y))$. 
Fig. 1 shows the doping dependence of $\Phi$ and $\Delta$ for $t'/t=0.35$ and $J=0.3$ at zero temperature. 
In the following  the lattice constant $a$ of the square lattice is used as length unit and $t$ as energy unit.
In the overdoped (OD) region,  
$\delta \geqslant \delta_c \sim 0.14$, 
$\Phi$ is zero while in the underdoped (UD) region, $\delta \leqslant \delta_c$,  $\Phi$ is nonzero and coexists with $\Delta$. The two 
order parameters compete each other leading to the decay of $\Delta$ with underdoping.\cite{cappelluti99,zeyher02} 
Inset (a) shows the temperature dependence of $\Phi$ for $\delta=0.10$ discarding superconductivity, the instability occurs when the 
flux susceptibility\cite{greco08} ($\chi_{flux}$) diverges   at 
$T_{FP} \sim0.018$ for momentum $q\sim (\pi,\pi)$. 
$\chi_{flux}({\bf q},\omega)=(\frac{\delta}{2})^2 [(8/J) r^2-\Pi({\bf q},\omega)]^{-1}$
where  $\Pi({\bf q},\omega)$ 
is the electronic polarizability calculated with a form factor 
$\gamma({\bf q},{\bf k})=2 r (sin(k_x-q_x/2)-sin(k_y-q_y/2))$; note that for ${\bf q} \sim (\pi,\pi)$ the form factor $\gamma({\bf q},{\bf k})$ transforms 
into $\sim (cos(k_x)-cos(k_y))$ which indicates the $d$-wave character of the FP.
From the dynamical point of view, with decreasing temperature from above,  a $d$-wave flux mode  with momentum $q=(\pi,\pi)$ 
softens,  increases its spectral weight and  reaches $\omega=0$ at $T_{FP}$ (inset (d)).  Below $T_{FP}$ the translational symmetry is broken. 
Inset (b) shows the evolution of the Fermi surface with decreasing temperature. In the normal state 
above $T_{FP}$  the FS is large and $(\pi,\pi)$-centered. Below $T_{FP}$, $\Phi$  opens near 
$(\pi,0)$ and the new FS consists of four small hole pockets near nodal direction. With 
decreasing $T$, $\Phi$ shows a fast increasing (inset (a)) and the side of hole  pockets is strongly diminished. 
However, the  spectral weight intensity of the outer part of the pocket is low,\cite{chakravarty03}
resembling  the presence of arcs (inset (c)).
Although the mean field FP scenario shows features (phase diagram, two competing gaps, pockets, arcs, and other ones \cite{zeyher02,greco04,chakravarty08})
which may be assimilated with those observed in cuprates, there are drawbacks to be addressed. 
While the FP scenario predicts a true phase transition at $T_{FP}$, some experiments show a smooth crossover.\cite{tallon01,kanigel06,kanigel07}
In addition, FP predicts well defined QP peaks above and below $T_{FP}$ everywhere on the Brillouin zone (BZ) while ARPES shows coherent 
peaks only near nodal direction.\cite{kim98,damascelli03} Near the anti-node  the QP peak is always broad, \cite{kim98,damascelli03} presenting 
a PG which seems to be filling up but not closing with increasing $T$\cite{norman98b,kanigel06,kanigel07} giving the impression of a smooth crossover 
of the spectral properties. In what follows it will be shown that this controversy can be solved after 
including self-energy effects. 

For discussing self-energy effects a large-$N$ ($N$ is the number of spin components) approach 
is useful\cite{foussats04} since  it is possible to identify the most relevant  fluctuations above mean field. As discussed in Ref.[\onlinecite{greco08}] the dominant 
contribution at low doping and low energy (thus a candidate for describing the 
PG) is given by the coupling between quasi-particles and FP fluctuations. 
The ${\cal O} (1/N)$ scattering rate reads
\begin{eqnarray}\label{eq:SigmaIm0}
Im \, \Sigma({\mathbf{k}},\omega)&=&-\frac{1}{N_{s}}\sum_{{\mathbf{q}}} \gamma^2({\bf q},{\bf {k}}) Im \chi_{flux}({\bf q},\omega-\epsilon_{{k-q}}) \nonumber\\
&& \hspace{-1cm}\times \left[n_{F}(-\epsilon_{{k-q}}) + n_{B}(\omega-\epsilon_{{k-q}})\right] \nonumber
\end{eqnarray}
\noindent where $n_B$  is the Bose  factor. Using the self-energy $\Sigma({\bf k},\omega)$ the spectral function $A({\bf k},\omega)$ is calculated as usual. Note that in ${\cal O} (1/N)$ the mean field transition temperature $T_{FP}$ is not renormalized by fluctuations. 
Finally, it is important to point out that present calculation is also a way for discussing the role of  self-energy 
corrections in the phenomenological   
$d$-CDW theory.
\begin{figure}
\begin{center}
\setlength{\unitlength}{1cm}
\includegraphics[width=8.8cm,angle=0.]{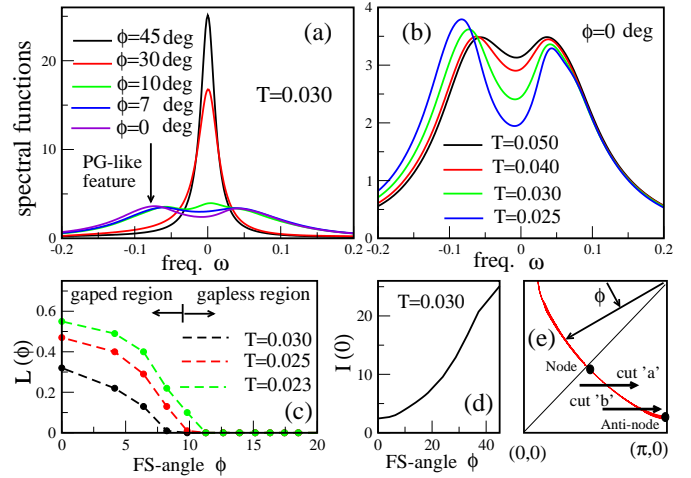}
\end{center}
\caption{ (color online) (a) Spectral functions at $T=0.030$ for several Fermi momenta  
from the node ($\phi=45$ deg) to the anti-node ($\phi=0$ deg) for $\delta=0.10$. 
(b) Spectral functions at the anti-node for several temperatures for $\delta=0.10$. 
c) Loss of the low-energy spectral weight $L(\phi)$ along the FS for different temperatures and  $\delta=0.10$. 
The length of the gaped (gapless) region 
decreases (increases) with increasing $T$.
(d) Intensity of the spectral functions at $\omega=0$ ($I(0)$) along the FS for $T=0.030$ and $\delta=0.10$. (e) Definition of the FS-angle $\phi$. The nodal and anti-nodal Fermi momenta  are indicated. The arrows show cut 'a' and cut 'b' along which the spectral functions of Fig. 4 were calculated. 
$\omega$ is measured with respect to the Fermi energy. A broadening $\eta=0.01$ was used in the calculation of the spectral functions. 
 }\label{fig1}
\end{figure}
\begin{figure}
\begin{center}
\setlength{\unitlength}{1cm}
\includegraphics[width=8.8cm,angle=0.]{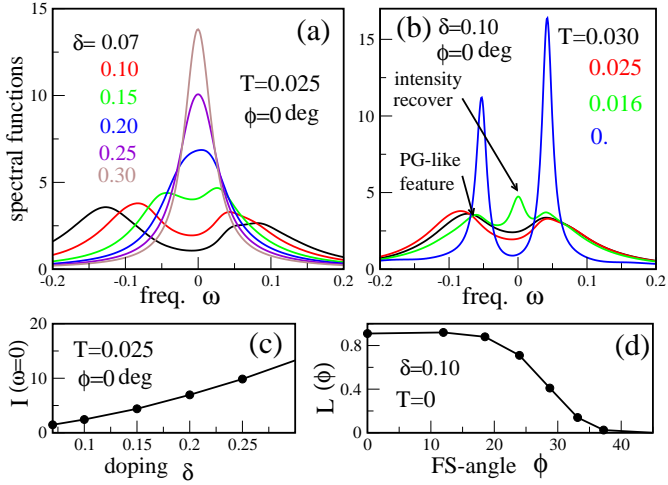}
\end{center}
\caption{(color online) (a) Spectral functions for $T=0.025$ at the anti-node for several dopings.
(b) Spectral functions at the anti-node for  $\delta=0.10$. For $T=0.030$ and $T=0.025$ 
the broad PG-like features are due to self-energy effects while for $T=0$ the true gap $\Phi$ dominates and the  peak at the PG becomes 
sharp. 
In  the transitional region at $T=0.016  \lesssim T_{FP}$ the broad PG-like feature at $\omega \sim -0.07$, due to self-energy effects, remains and 
a recover of the intensity at $\omega=0$ is observed.
(c) Intensity at $\omega=0$ as a function of doping at the anti-node for  $T=0.025$.  
(d) Loss of the low-energy spectral weight at  $T=0$ for  $\delta=0.10$. 
 }\label{fig1}
\end{figure}

Present theoretical results will be presented in the standard form for an easier comparison with ARPES experiments.
Fig. 2(a)  shows $A({\bf k},\omega)$ for $\delta=0.10$ (Ref.[\onlinecite{doping}])  and   $T=0.030 > T_{FP}$ for different momenta, 
on the normal state FS, labeled by the FS-angle $\phi$ (see Fig. 2(e)).
A gapless spectrum with maximun at $\omega=0$ is observed near the node ($\phi=45$ deg). Moving towards the anti-node ($\phi=0$) 
the spectrum become broad, losses intensity at $\omega=0$ and develops a PG-like feature, suggesting the presence of arcs.
Decreasing temperature approaching $T_{FP}$ the gaped region near the anti-node expands, e.i., the length of the arc decreases.
Note that the arcs developed at $T>T_{FP}$ do not imply a breaking of the translational symmetry.
The behavior of the intensity at $\omega=0$ ($I(0)$) along  the FS (Fig. 2(d)) indicates  also the loss of 
QP coherence near the anti-node (see Fig. 2(b) in Ref. [\onlinecite{norman07}] for comparison with experiment).  
These PG-like features occur above but near $T_{FP}$ and are  associated to
self-energy effects proving the proximity to the FP instability via the coupling between quasi-particles and the $d$-wave soft flux modes. As discussed above,
the soft mode has momentum $q\sim(\pi,\pi)$ allowing the scattering between the quasi-particles  near the anti-nodes. 
Adopting the accepted value $t=0.4eV$ the 
temperature range and the PG have the same order of magnitude than those in the experiment.
\begin{figure}
\begin{center}
\setlength{\unitlength}{1cm}
\includegraphics[width=8.8cm,angle=0.]{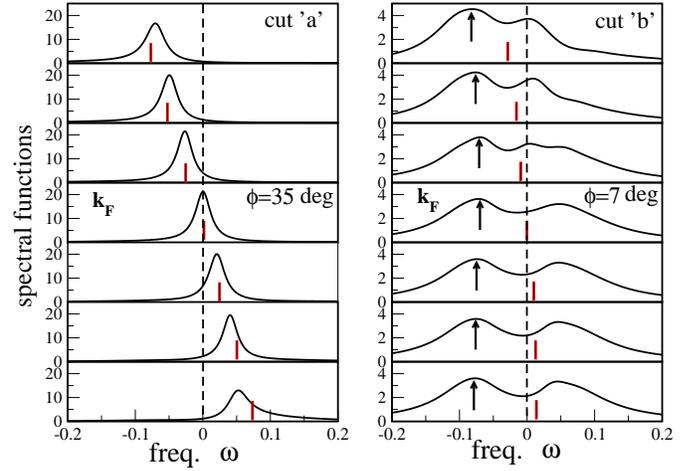}
\end{center}
\caption{(color online) 
Spectral functions for $\delta=0.10$ along the cut 'a' (left panel) and cut 'b' (right panel) at $T=0.025$. 
Cut 'a' (Cut 'b') passes trough the gapless (gaped) Fermi momentum at $\phi=35$ deg ($\phi=7$ deg). 
Vertical dashed line indicates the Fermi level. Red marks show 
the location of the unrenormalized QP peaks. 
}\label{fig1}
\end{figure}

For comparison with ARPES results \cite{kanigel07} Fig. 2(c) shows the loss of the low-energy spectral weight
($L(\phi)=1-I(0)/I(\Delta_\phi)$) along the FS for different temperatures. ($I(\Delta_\phi)$ is the intensity at the PG as 
exemplified  by the arrow  in Fig. 2(a)).  
In agreement with the experiment (see Fig. 3 of Ref. [\onlinecite{kanigel07}]) the arc length expands with increasing $T$. 
Note also that $L(\phi=0)$ decreases with increasing temperature.\cite{kanigel07}  
In addition, the temperature behavior of the spectral functions 
at the anti-node indicates that the 
PG-like feature is filling but not closing (Fig. 2(b)). This behavior leads to a PG rather temperature independent in qualitative agreement with Refs. [\onlinecite{norman98b,kanigel06,kanigel07}]. 
In the experiment of Ref. [\onlinecite{kanigel07}] this behavior seems to continue in the superconducting state, thus the existence of only one gap 
was assumed as favoring the interpretation of preformed pairs. 
It is worth to mention that on a two-gap scenario the explanation  of this behavior is not straightforward (see for instance Ref.[\onlinecite{mark}]). 
However, in several ARPES experiments\cite{ma08,yoshida09,kondo09} two gaps are  detected, making doubtful the interpretation of Ref. [\onlinecite{kanigel07}]. In spite of this discrepancy is presently unclear, in the 
present paper it is shown that some ARPES data can be described in the context of the two-gap scenario.

Next, the dependence with doping is discussed. Fig. 3(a) shows the spectral functions at the anti-node for different  dopings at $T=0.025$.  
With overdoping the PG-like feature closes and, simultaneously,  the intensity at $\omega=0$ increases (see also Fig. 3(c)), emerging the 
QP peak  as in the experiment. \cite{kim98,kaminski05} 
Note that from present approach the PG-like features are expected, in the normal state, even 
for $\delta \gtrsim \delta_c$. This fact indicates that there is no contradiction between the existence of PG features in nearly overdoped samples\cite{damascelli03} and theories, like $d$-CDW or FP, which suggest the existence of a quantum critical point near optimal 
doping.

In the following  it will be shown that recent ARPES results \cite{kanigel08} (see also Ref. [\onlinecite{shi08}]), 
interpreted as evidence for preformed pairs above $T_c$ due to the apparent Bogoliubov-like dispersion 
of quasiparticles near the anti-node, are also consistent with a theory for two competing order parameters. 
Fig. 4 shows spectral functions for $\delta=0.10$ 
at $T=0.025$ along cut 'a' (left panel ) and 
cut 'b' (right panel) which are parallel to the $(0,0)-(\pi,0)$ direction (Fig. 2(e)).  As in Fig. 3 of Ref. [\onlinecite{kanigel08}] the peaks are better defined in the arc, or gapless region (cut 'a'), than in the gaped region (cut 'b')
where only broad features are seen. While for cut 'a' the low energy peak approaches $k_F$, for cut 'b' the broad feature 
(marked with arrow in right panel)  
back bends slightly. 
According to present theory the reason for this behavior relays on a different fact to that discussed in Ref. [\onlinecite{kanigel08}].
Indeed,  the broad feature  along cut 'b' 
is due to self-energy effects coming from the interaction between quasi-particles  and the soft modes in the proximity to the $d$-CDW or FP instability.

Now, results for $T < T_{FP}$ are discussed.
For $T \ll T_{FP}$ a large gap $\Phi$  occurs in the electronic spectrum what affects  $\chi_{flux}$ and washes the 
effect of the low energy $d$-CDW soft mode on the self-energy. 
In this case the expected situation is close to that predicted for the pure $d$-CDW picture (see Fig. 3(b) for $T=0$) and 
pockets would be observed.\cite{add} 
On the other hand, the peak at the PG becomes much sharper than those observed at $T > T_{FP}$ as 
can be seen by comparing the result for $T=0$ with those for $T=0.030$ and $T=0.025$. 
In addition, at $T=0$, the loss of the low-energy spectral weight $L(\phi)$ (Fig. 3(d)) follows the typical behavior expected for a $d$-wave gap.  
There is a range of temperatures below but near  $T_{FP}$ where $\Phi$ is small (Fig. 1(a)) and its value competes with the energy of the 
soft mode. In this region a recovery of the intensity at $\omega=0$ is observed together with the broad PG-like feature 
that appears at $T > T_{FP}$ (see Fig. 3(b) for $T=0.016$).
Note that the range of temperatures which marks the transition from PG-like features due to self-energy effects for $T > T_{FP}$ towards the 
situation mainly dominated by the true gap $\Phi$ for $T \ll T_{FP}$ is small. 
Searching in the above findings could  contribute to the discussion about the dichotomy between: a) Fermi arcs\cite{kanigel06,kanigel07,hossain08} and Fermi pockets,\cite{chang08,doiron07} b) smooth crossover\cite{kanigel06,kanigel07,tallon01} and abrupt transition
\cite{fauque06,mook08,leridon09} and c)  one gap\cite{kanigel07} and two gaps.\cite{ma08,yoshida09}

In summary, it was shown that recent ARPES experiments do not necessarily imply the existence of preformed pairs above $T_c$. These experiments 
can also be described in the context of the two competing order parameters if self-energy effects are take into account in the proximity 
of the opening of a normal state gap like in the case of $d$-CDW or the flux phase. Finally, present paper supports theories invoking two 
competing gaps as the explanation for the low doping properties of cuprates. 

The author thanks to M. Bejas, J.C. Campuzano, A. Foussats, A. Muramatsu, H. Parent, and R. Zeyher for valuable discussions.


\begin{thebibliography}{10}
\bibitem{doping} The doping value $\delta=0.10$ is similar to the corresponding doping of the underdoped samples studied 
in Refs. [\onlinecite{kanigel06,kanigel07,kanigel08}].
\bibitem{add} Since the present paper is focused in the pseudogap regime, superconductivity was discarded. 
Due to the fast increasing of $\Phi$ with decreasing $T$ results for $T=0$ are expected for  $T > T_c$.
\end{thebibliography}
\end{document}